\documentstyle[twoside,psfig,amsfonts]{article}
\catcode`\@=11
\long\def\@makefntext#1{
\protect\noindent \hbox to 3.2pt {\hskip-.9pt  
$^{{\eightrm\@thefnmark}}$\hfil}#1\hfill}		%CAN BE USED 

\def\thefootnote{\fnsymbol{footnote}}
\def\@makefnmark{\hbox to 0pt{$^{\@thefnmark}$\hss}}	%ORIGINAL 
	
\def\ps@myheadings{\let\@mkboth\@gobbletwo
\def\@oddhead{\hbox{}
\rightmark\hfil\eightrm\thepage}   
\def\@oddfoot{}\def\@evenhead{\eightrm\thepage\hfil
\leftmark\hbox{}}\def\@evenfoot{}
\def\sectionmark##1{}\def\subsectionmark##1{}}

\oddsidemargin=\evensidemargin
\renewcommand{\thefootnote}{\fnsymbol{footnote}}

\newcounter{sectionc}\newcounter{subsectionc}\newcounter{subsubsectionc}
\renewcommand{\section}[1] {\vspace{12pt}\addtocounter{sectionc}{1} 
\setcounter{subsectionc}{0}\setcounter{subsubsectionc}{0}\noindent 
	{\tenbf\thesectionc. #1}\par\vspace{5pt}}
\renewcommand{\subsection}[1] {\vspace{12pt}\addtocounter{subsectionc}{1} 
	\setcounter{subsubsectionc}{0}\noindent 
	{\bf\thesectionc.\thesubsectionc. {\kern1pt \bfit #1}}\par\vspace{5pt}}
\renewcommand{\subsubsection}[1] {\vspace{12pt}\addtocounter{subsubsectionc}{1}
	\noindent{\tenrm\thesectionc.\thesubsectionc.\thesubsubsectionc.
	{\kern1pt \tenit #1}}\par\vspace{5pt}}
\newcommand{\nonumsection}[1] {\vspace{12pt}\noindent{\tenbf #1}
	\par\vspace{5pt}}

\newcounter{appendixc}
\newcounter{subappendixc}[appendixc]
\newcounter{subsubappendixc}[subappendixc]
\renewcommand{\thesubappendixc}{\Alph{appendixc}.\arabic{subappendixc}}
\renewcommand{\thesubsubappendixc}
	{\Alph{appendixc}.\arabic{subappendixc}.\arabic{subsubappendixc}}

\renewcommand{\appendix}[1] {\vspace{12pt}
        \refstepcounter{appendixc}
        \setcounter{figure}{0}
        \setcounter{table}{0}
        \setcounter{lemma}{0}
        \setcounter{theorem}{0}
        \setcounter{corollary}{0}
        \setcounter{definition}{0}
        \setcounter{equation}{0}
        \renewcommand{\thefigure}{\Alph{appendixc}.\arabic{figure}}
        \renewcommand{\thetable}{\Alph{appendixc}.\arabic{table}}
        \renewcommand{\theappendixc}{\Alph{appendixc}}
        \renewcommand{\thelemma}{\Alph{appendixc}.\arabic{lemma}}
        \renewcommand{\thetheorem}{\Alph{appendixc}.\arabic{theorem}}
        \renewcommand{\thedefinition}{\Alph{appendixc}.\arabic{definition}}
        \renewcommand{\thecorollary}{\Alph{appendixc}.\arabic{corollary}}
        \renewcommand{\theequation}{\Alph{appendixc}.\arabic{equation}}
        \noindent{\tenbf Appendix \theappendixc #1}\par\vspace{5pt}}
\newcommand{\subappendix}[1] {\vspace{12pt}
        \refstepcounter{subappendixc}
        \noindent{\bf Appendix \thesubappendixc. {\kern1pt \bfit #1}}
	\par\vspace{5pt}}
\newcommand{\subsubappendix}[1] {\vspace{12pt}
        \refstepcounter{subsubappendixc}
        \noindent{\rm Appendix \thesubsubappendixc. {\kern1pt \tenit #1}}
	\par\vspace{5pt}}

\newcommand{\textlineskip}{\baselineskip=13pt}
\newcommand{\smalllineskip}{\baselineskip=10pt}

\def\eightcirc{
\begin{picture}(0,0)
\put(4.4,1.8){\circle{6.5}}
\end{picture}}
\def\eightcopyright{\eightcirc\kern2.7pt\hbox{\eightrm c}}

\def\abstracts#1#2#3{{
	\centering{\begin{minipage}{4.5in}\footnotesize\baselineskip=10pt
	\parindent=0pt #1\par 
	\parindent=15pt #2\par
	\parindent=15pt #3
	\end{minipage}}\par}} 

\newcommand{\bibit}{\nineit}

\renewenvironment{thebibliography}[1]
	{\frenchspacing
	 \ninerm\baselineskip=11pt
	 \begin{list}{\arabic{enumi}.}
	{\usecounter{enumi}\setlength{\parsep}{0pt}
\setlength{\leftmargin 17pt}{\rightmargin 0pt}   %FOR 10--99 ITEMS
	 \setlength{\itemsep}{0pt} \settowidth
	{\labelwidth}{#1.}\sloppy}}{\end{list}}

\newcounter{itemlistc}
\newcounter{romanlistc}
\newcounter{alphlistc}
\newcounter{arabiclistc}

\newcommand{\fcaption}[1]{
        \refstepcounter{figure}
        \setbox\@tempboxa = \hbox{\footnotesize Fig.~\thefigure. #1}
        \ifdim \wd\@tempboxa > 5in
           {\begin{center}
        \parbox{5in}{\footnotesize\smalllineskip Fig.~\thefigure. #1}
            \end{center}}
        \else
             {\begin{center}
             {\footnotesize Fig.~\thefigure. #1}
              \end{center}}
        \fi}

\newcommand{\tcaption}[1]{
        \refstepcounter{table}
        \setbox\@tempboxa = \hbox{\footnotesize Table~\thetable. #1}
        \ifdim \wd\@tempboxa > 5in
           {\begin{center}
        \parbox{5in}{\footnotesize\smalllineskip Table~\thetable. #1}
            \end{center}}
        \else
             {\begin{center}
             {\footnotesize Table~\thetable. #1}
              \end{center}}
        \fi}

\def\@citex[#1]#2{\if@filesw\immediate\write\@auxout
	{\string\citation{#2}}\fi
\def\@citea{}\@cite{\@for\@citeb:=#2\do
	{\@citea\def\@citea{,}\@ifundefined
	{b@\@citeb}{{\bf ?}\@warning
	{Citation `\@citeb' on page \thepage \space undefined}}
	{\csname b@\@citeb\endcsname}}}{#1}}

\newif\if@cghi
\def\cite{\@cghitrue\@ifnextchar [{\@tempswatrue
	\@citex}{\@tempswafalse\@citex[]}}
\def\citelow{\@cghifalse\@ifnextchar [{\@tempswatrue
	\@citex}{\@tempswafalse\@citex[]}}
\def\@cite#1#2{{$\null^{#1}$\if@tempswa\typeout
	{IJCGA warning: optional citation argument 
	ignored: `#2'} \fi}}

\def\pmb#1{\setbox0=\hbox{#1}
	\kern-.025em\copy0\kern-\wd0
	\kern.05em\copy0\kern-\wd0
	\kern-.025em\raise.0433em\box0}

\def\fnt#1#2{\footnotetext{\kern-.3em
	{$^{\mbox{\scriptsize #1}}$}{#2}}}

\def\thefootnote{\fnsymbol{footnote}}
\def\@makefnmark{\hbox to 0pt{$^{\@thefnmark}$\hss}}	%ORIGINAL 
	
\def\ps@myheadings{%
    \def\@evenhead{\slshape\leftmark\hfil}%       %EVEN PAGE
    \def\@oddhead{\hfil{\slshape\rightmark}}%     %ODD PAGE
    \let\@mkboth\@gobbletwo
    \let\sectionmark\@gobble
    \let\subsectionmark\@gobble
    }
\font\tenrm=cmr10
\font\tenit=cmti10 
\font\tenbf=cmbx10
\font\bfit=cmbxti10 at 10pt
\font\ninerm=cmr9
\font\nineit=cmti9

\font\eightrm=cmr8

\def\qed{\hbox{${\vcenter{\vbox{			%HOLLOW SQUARE
   \hrule height 0.4pt\hbox{\vrule width 0.4pt height 6pt
   \kern5pt\vrule width 0.4pt}\hrule height 0.4pt}}}$}}

\renewcommand{\thefootnote}{\fnsymbol{footnote}}  %USE SYMBOLIC FOOTNOTE

\def\lp{\stackrel{\leftarrow}{\partial}}
\def\rp{\stackrel{\rightarrow}{\partial}}
\def\be{\begin{eqnarray}}
\def\ee{\end{eqnarray}}
\def\*{\star}
\def\bigstar{\;\bigcirc\kern-0.9em\star\;} 
\begin{document}
\thispagestyle{empty}
\setlength{\textheight}{7.7truein}  %for 2nd page onwards
\markboth{\protect{\footnotesize\it 
C K Zachos }}{\protect{\footnotesize\it 
hep-th/0110114 \qquad \qquad \qquad \qquad Deformation Quantization}} 
\normalsize\textlineskip
\setcounter{page}{1}
          \begin{flushleft}  
          ANL-HEP-PR-01-095 
          \end{flushleft}
%\copyrightheading{}		%{Vol.~0, No.~0 (2000) 000--000}
\vspace*{0.68truein}
%\fpage{1}
\centerline{\bf 
DEFORMATION QUANTIZATION:} \vspace*{0.035truein}
\centerline{\bf QUANTUM MECHANICS LIVES AND WORKS IN PHASE-SPACE}  
\vspace*{0.37truein}
\centerline{\footnotesize COSMAS ZACHOS}
\baselineskip=12pt
\centerline{\footnotesize\it High Energy Physics Division,
Argonne National Laboratory}
\baselineskip=10pt
\centerline{\footnotesize\it 
 Argonne, IL 60439-4815, USA }
\baselineskip=10pt
\centerline{\footnotesize\it  E-mail: zachos@anl.gov }
\vspace*{0.225truein}
%\publisher{(received date)}{(revised date)}

\vspace*{0.21truein}
\abstracts{ Wigner's quasi-probability distribution function in phase-space
is a special (Weyl) representation of the density matrix. It has been useful 
in describing quantum transport in quantum optics; nuclear physics; 
decoherence (eg, quantum computing); quantum chaos; 
``Welcher Weg" discussions; semiclassical limits. It is also of importance 
in signal processing.
 Nevertheless, a remarkable aspect of its internal logic, pioneered by 
Moyal, has only emerged in the last quarter-century:
It furnishes a third, alternative, formulation of Quantum Mechanics, 
independent of the conventional Hilbert Space, or Path Integral formulations.
In this logically complete and self-standing formulation, one need not choose 
sides---coordinate or momentum space. It works in full phase-space, 
accommodating the uncertainty principle. This is an introductory overview of 
the formulation with simple illustrations.} {}{}
\setcounter{footnote}{0}
\renewcommand{\thefootnote}{\alph{footnote}}

\vspace*{3pt}\textlineskip	
\section{Introduction}	
\vspace*{-0.5pt}
\noindent
There are three logically autonomous alternative  paths to quantization.
The first is the standard one utilizing operators in Hilbert space,
developed by Heisenberg, Schr\"{o}dinger, Dirac, and others in the Twenties of 
the past century. The second one relies on Path Integrals, and was 
conceived by Dirac\cite{pamd} and constructed by Feynman. 

The last one (the bronze medal!) is the Phase-Space formulation, based on 
Wigner's (1932) 
quasi-distribution function\cite{wigner} and Weyl's (1927) 
correspondence\cite{weyl} between quantum-mechanical 
operators and ordinary c-number 
phase-space functions.
The crucial composition structure of these functions, which relies 
on the $\*$-product, was 
fully understood by Groenewold (1946)\cite{groen}, who, together 
with Moyal (1949)\cite{moyal}, pulled the entire formulation together. 
Still, insights on interpretation and 
a full appreciation of its conceptual autonomy took some time to mature with 
the work of Takabayasi\cite{takabayasi}, Baker\cite{baker}, Fairlie 
\cite{dbf}, and others.
(A brief guide to some landmark papers is provided in the Appendix.)

This complete formulation is based on the Wigner Function (WF), which is a 
quasi-probability distribution function in phase-space, 
\be
{f(x,p)={1\over 2\pi}\int\! dy~\psi ^* 
(x-{\hbar\over 2} y )~e^{-iyp} \psi(x+{\hbar\over2} y)}.   
\ee
It is a special representation of the density matrix (in the Weyl 
correspondence, as detailed in Section 8). Alternatively, it is a 
generating function for all spatial autocorrelation functions of a given 
quantum-mechanical wavefunction $\psi(x)$. 

There are several outstanding reviews on the 
subject\cite{reviews,berry,bayen}. 
Nevertheless, the central conceit of this review is that the above input 
wavefunctions  may ultimately be forfeited, since the WFs are determined, 
in principle, as the solutions of suitable (celebrated) functional
equations. Connections to the Hilbert space operator formulation of quantum
mechanics may thus be ignored, in principle---even though they are 
provided in Section 8 for pedagogy and confirmation of the formulations' 
equivalence. One might then envision an imaginary planet on which this 
formulation of Quantum Mechanics preceded the conventional formulation, 
and its own techniques and methods arose independently, perhaps out of 
generalizations of classical mechanics and statistical mechanics.

It is not only wavefunctions that are missing in this formulation. Beyond 
an all-important (noncommutative, associative, pseudodifferential) operation, 
the $\*$-product, which encodes the entire quantum mechanical action, 
there are no operators. Observables and transition amplitudes 
are  phase-space integrals of c-number functions (which compose through the 
$\*$-product), weighted by the WF, as in statistical mechanics. Consequently,
even though the WF is not positive-semidefinite (it can, and usually does go
negative in parts of phase-space), the computation of observables and the 
associated concepts are evocative of classical probability theory.

This formulation of Quantum Mechanics is useful in describing 
{\em quantum} transport processes in phase space, of importance in  
quantum optics, nuclear physics, condensed matter, 
and the study of semiclassical limits of mesoscopic 
systems\cite{semiclassical} and the transition 
to classical statistical mechanics\cite{sundryphys}. 
It is the natural language to study 
quantum chaos and decoherence\cite{decoherence} 
(of utility in, eg, quantum computing), and provides intuition in QM 
interference such as in ``Welcher Weg" problems\cite{wiseman},
probability flows as negative probability backflows\cite{negprop}, and 
measurements of atomic systems\cite{exper}. 
The intriguing mathematical structure 
of the formulation is of relevance to Lie Algebras\cite{fz}, 
and has recently been  retrofitted into M-theory advances linked to 
noncommutative geometry\cite{natied}, and matrix models\cite{taylor}, which 
apply spacetime uncertainty  principles\cite{noncom} reliant 
on the $\*$-product. (Transverse spatial dimensions
act formally as momenta, and, analogously to quantum mechanics, their
uncertainty is increased or decreased inversely to the uncertainty of a given
direction.)

As a significant aside, the WF has extensive practical applications in 
signal processing and engineering (time-frequency analysis), since 
time and energy (frequency) constitute a pair of Fourier-conjugate variables 
just like the ${\mathfrak x}$ and ${\mathfrak  p}$ of phase space\footnote{Thus,
 time varying signals are best represented in a WF as time 
varying spectrograms, analogously to a music score, ie the changing 
distribution of 
frequencies is monitored in time\cite{score}: even though the 
description is constrained and redundant, it gives an intuitive 
picture of the signal that a mere time profile or frequency spectrogram 
fails to convey. Applications abound in acoustics, speech analysis, 
seismic data analysis, internal combustion engine-knocking analysis, 
failing helicopter component 
vibrations, etc.}. 

For simplicity, the formulation will be mostly illustrated for one coordinate 
and its conjugate momentum, but generalization to arbitrary-sized phase 
spaces is straightforward\cite{dodonov}, including infinite-dimensional 
ones, namely scalar field theory\cite{dito,lesche,nachbag,base}.

\section{The Wigner Function}
\noindent
As already indicated, the quasi-probability measure in phase space is the WF,
\be
{f(x,p)={1\over 2\pi}\int\! 
dy~\psi^* (x-{\hbar\over2} y )~e^{-iyp} 
\psi(x+{\hbar\over2} y)}.   \label{wiggie}
\ee
It is obviously normalized, $\int dp dx f(x,p)=1$. In the classical limit,
$\hbar\rightarrow 0$, it would reduce to the probability density in 
coordinate space $x$, usually highly localized, multiplied by 
$\delta$-functions in momentum:
the classical limit is ``spiky"! This expression has more $x-p$ symmetry than 
is apparent, as Fourier transformation to momentum-space wavefunctions 
yields a completely symmetric expression with the roles of $x$ and $p$ reversed,
and, upon rescaling of the arguments $x$ and $p$, a symmetric classical limit.

It is also manifestly Real\footnote{In one space dimension, by virtue of 
non-degeneracy, it turns out to be $p$-even, but this is not a property used
here.}. It is also constrained by the Schwarz Inequality to be  bounded,
$-\frac 2 h \leq f(x,p)  \leq \frac 2 h $. Again, this bound disappears in the 
spiky classical limit.

$p$- or $x$-projection leads to  marginal (bona-fide) probability densities: 
a spacelike shadow $\int dp f(x,p)=\rho(x)$, {\em or else} a 
momentum-space shadow $\int dx f(x,p)=\sigma (p)$, respectively.
Either is a (bona-fide) probability density, being positive semidefinite. 
But neither can be conditioned on the other, as the uncertainty principle is 
fighting back:  The WF $f(x,p)$ itself can, and most often does go negative
in some areas of phase-space\cite {wigner,reviews}, as is illustrated 
below, a hallmark of QM interference in this language. (In fact, the only 
pure state WF which is non-negative is the Gaussian\cite{hudson}.)

The counter-intuitive ``negative probability"  aspects of this 
quasi-probability distribution have been explored and 
interpreted\cite{bartlett,feynman,negprop}, and negative probability flows 
amount to legitimate probability backflows in interesting 
settings\cite{negprop}.  Nevertheless, the WF for atomic systems can still be 
measured in the laboratory, albeit indirectly\cite{exper}. 

Smoothing $f$ by a filter of size larger than $\hbar$ (eg, convolving with a
phase-space Gaussian) results in a positive-semidefinite 
function, ie it may be thought to have been coarsened to a classical 
distribution\cite{cartwright}\footnote{This one is called the Husimi 
distribution\cite{takahashi}, and sometimes information scientists 
examine it on account of its non-negative feature. Nevertheless, it comes 
with a heavy price, as it needs to be ``dressed" back to the WF for 
all practical purposes when expectation values are computed with it,
ie it does not serve as an immediate quasi-probability distribution with no 
other measure\cite{takahashi}. The negative feature of the WF is, in the 
last analysis, an asset, and not a liability, and provides an efficient 
description of ``beats"\cite{score}.}.

Among real functions, the WFs comprise a rather small, highly  constrained,
set. When is a real function $f(x,p)$ a bona-fide Wigner function of the 
form (\ref{wiggie})? Evidently, when its Fourier transform 
``left-right" factorizes,
\be 
\tilde{f}(x,y)=\int dp ~e^{ipy} f(x,p) 
= g^{*}_L (x-\hbar y/2) ~g_R (x+\hbar y/2),
\ee
ie, 
\be
{\partial^2 \ln \tilde{f}\over \partial(x-\hbar y/2)~\partial(x+\hbar y/2)}
=0,   \label{test}
\ee
so $g_L=g_R$ from reality.      

Nevertheless, as indicated, the WF {\em is} a distribution function: 
it provides the integration measure in phase space to yield expectation 
values from phase-space c-number functions. Such functions are often
classical quantities, but, in general, are associated
to suitably ordered operators through 
Weyl's correspondence rule\cite{weyl}. Given an operator ordered 
in this prescription, 
\be {\mathfrak G(x,p)} =\frac{1}{(2\pi)^2}\int d\tau d\sigma 
dx dp ~g(x,p) \exp (i\tau ({ {\mathfrak  p}}-p)+i\sigma ({ {\mathfrak  x}}-x)),
\ee
the corresponding phase-space function $g(x,p)$ 
(the ``classical kernel of the operator") is obtained by 
\be
{ {\mathfrak  p}}\mapsto p , ~ { {\mathfrak  x}}\mapsto x .
\ee 
That operator's expectation value is then a ``phase-space average" 
\cite{groen,moyal}, 
\be
\langle { {\mathfrak G}} \rangle =\int\! dx dp~ f(x,p)~ g(x,p) .
\ee
The classical kernel $g(x,p)$ is often the unmodified classical expression,
such as a conventional hamiltonian, $H=\frac{p^2}{2m} + V(x)$, ie the 
transition from classical mechanics is straightforward ; but it contains 
$\hbar$ when there are quantum-mechanical ordering ambiguities, such as for the 
classical kernel of the square of the angular momentum 
${\mathfrak L}\cdot {\mathfrak L} $: this one contains 
$O(\hbar^2)$ terms introduced by the Weyl ordering, beyond the 
classical expression $L^2$. In such cases, the classical kernels may still
be produced without direct consideration of operators. 
A more detailed discussion of the correspondence is
provided in Section 8.

In this sense, expectation values of the physical observables specified by
classical kernels $g(x,p)$ are computed through integration with the WF,
in close analogy to classical probability theory, but for the 
non-positive-definiteness of the distribution function. This operation 
corresponds to tracing an operator with the density matrix (Sec 8).

\section{Solving for the Wigner Function}
\noindent
Given a specification of observables, the next step is to find the relevant 
WF for a given hamiltonian. 
Can this be done without solving for the Schr\"{o}dinger wavefunctions 
$\psi$, i.e. not using Schr\"{o}dinger's equation directly?
The functional equations which $f$ satisfies completely determine it. 

Firstly, its dynamical evolution is specified by Moyal's equation.
This is the extension of Liouville's theorem of classical mechanics,
for a classical hamiltonian $H(x,p)$, namely  
$\partial_t f+\{f,H\}=0$, 
to quantum mechanics, in this language\cite{wigner,moyal}: 
\be 
{{\partial f \over \partial t} 
={H \star f - f\star H \over i\hbar}}~, \label{evolution}
\ee
where the $\star$-product\cite{groen} is 
\be 
{\star \equiv e^{{i\hbar \over 2} (\lp_x \rp_p- \lp_p \rp_x )} } .  
\ee

The right hand side of (\ref{evolution}), dubbed the ``Moyal Bracket"  (the 
Weyl-correspondent of quantum commutators),  
is the essentially unique one-parameter ($\hbar$) associative 
deformation of the Poisson Brackets of classical mechanics\cite{vey}.
Expansion in $\hbar$ around 0 reveals that it consists of the Poisson 
Bracket corrected by terms $O(\hbar)$. The equation (\ref{evolution}) also
evokes Heisenberg's equation of motion for operators, except 
$H$ and $f$ here are classical functions, and it is the $\*$-product 
which introduces noncommutativity. This language, thus, makes the link 
between quantum commutators and Poisson Brackets more transparent. 

Since the $\*$-product involves exponentials of derivative operators, it 
may be evaluated in practice through translation 
of function arguments (``Bopp shifts"), 
\begin{equation}
f(x,p) \star g(x,p) = f\left( x+{i\hbar\over 2}\rp_p ,~ 
p-{i\hbar\over 2}\rp_x \right )~ g(x,p).
\end{equation}
The equivalent Fourier representation of the $\*$-product is\cite{baker}: 
\be 
f\star g &=&{1\over \hbar ^2 \pi^2}\int dp^{\prime}  
dp^{\prime\prime}  dx' dx''  ~f(x',p')~g(x'',p'') \nonumber \\
&=&   \exp \left({-2i\over \hbar} 
\left( p(x'-x'') + p'(x''-x)+p''(x-x') \right )\right) .  \label{fourier}
\ee 
An alternate integral representation of this product is\cite{neumann}
\begin{equation} 
f\star g=(\hbar \pi)^{-2} \int du dv dw dz ~f(x+u,p+v)~g(x+w,p+z)~ 
\exp \left({2i\over \hbar} 
\left( uz-vw \right) \right) ,  \label{shift}
\end{equation}
which readily displays noncommutativity and associativity. 

$\star$-multiplication of c-number phase-space functions is  
in complete isomorphism  to Hilbert-space operator multiplication\cite{groen},
\be 
{\mathfrak A B} =
\frac{1}{(2\pi)^2} \int d\tau d\sigma dx dp ~ (a \star b) ~ \exp 
(i\tau ({\mathfrak p}-p)+i\sigma ( {\mathfrak  x }-x)).
\ee
The cyclic phase-space trace is 
directly seen in the representation (\ref{shift}) to reduce to a plain 
product, if there is only one $\*$ involved,
\begin{equation}
\int\! dp dx ~ f\star g = \int\! dp dx ~ fg
=\int\! dp dx ~ g\star f ~ \label{Ndjambi}.
\end{equation}

Moyal's equation is necessary, but does not suffice to specify the WF
for a system. In the conventional formulation of quantum mechanics,
systematic solution of time-dependent equations is usually predicated on the 
spectrum of stationary ones. Time-independent pure-state Wigner 
functions $\star$-commute with $H$, but clearly not every function 
$\star$-commuting with $H$ can be a bona-fide WF (eg, any $\star$-function 
of $H$ will $\star$-commute with $H$).

Static WFs obey more powerful functional $\star$-genvalue equations\cite{dbf} 
(also see \cite{starothers}),
\be 
H(x,p)\star f(x,p)& =& H\left (x+{i\hbar\over 2} \rp_p ~,~ 
p-{i\hbar\over 2} \rp_x\right ) ~ f(x,p) \nonumber \\
&=& f(x,p)   \star H(x,p)= E ~f(x,p)~, \label{stargenvalue}
\ee
where $E$ is the energy eigenvalue of ${\mathfrak H} \psi = E \psi$.
These amount to a complete characterization of the WFs\cite{cfz}.
 
\vspace*{12pt} \noindent {\bf Lemma:}
{\em For real functions $f(x,p)$, the Wigner form (\ref{wiggie}) 
for pure static eigenstates is equivalent to 
compliance with the $\star$-genvalue equations (\ref{stargenvalue}) 
($\Re $ and $\Im$ parts)}. 
%\vspace*{9pt} 
\noindent {\bf Proof:} 
\begin{eqnarray}             
&& H(x,p)\star f(x,p)= \nonumber \\
&=& {1\over 2\pi} \left( (p-i{\hbar\over2} \rp_x)^2/2m 
+V(x) \right) \int\! dy~ e^{-iy(p+i{\hbar\over2} \lp_x)}
\psi^*(x-{\hbar\over2} y) ~\psi(x+{\hbar\over2} y) \nonumber  \\ 
&=& {1\over 2\pi} \int\! dy~ \left( (p-i{\hbar\over2} \rp_x)^2/2m 
+V(x+{\hbar\over2}y)\right)   e^{-iyp}\psi^*(x-{\hbar\over2} y)~
\psi(x+{\hbar\over2} y) \nonumber \\
&=& {1\over 2\pi} \int\! dy ~e^{-iyp} 
\left( (i\rp_y +i{\hbar\over2} \rp_x)^2/2m + V(x+{\hbar\over2}y)\right)  
\psi^*(x-{\hbar\over2} y) ~\psi(x+{\hbar\over2} y)  \nonumber 
\\ &=& {1\over 2\pi} \int\! dy ~e^{-iyp} \psi^*(x-{\hbar\over2} y)~
E~ \psi(x+{\hbar\over2} y)  \nonumber \\
&=& {E ~f(x,p)}.
\end{eqnarray}
Action of the effective differential operators on 
$\psi^*$ turns out to be null. 

Symmetrically,    \pagebreak 
\begin{eqnarray}             
&\phantom{b}& f\star H = \nonumber \\
&=& {1\over 2\pi} \int\! dy ~e^{-iyp} 
\left( - {1\over 2m} (\rp_y -{\hbar\over2} \rp_x)^2 
+ V(x-{\hbar\over2}y)\right)  
\psi^*(x-{\hbar\over2} y) ~\psi(x+{\hbar\over2} y) \nonumber \\ 
&=& E~ f(x,p), 
\end{eqnarray}
where the action on $\psi$ is now trivial. 
Conversely, the pair of $\star $-eigenvalue equations 
dictate, for $f(x,p)=\int\!  dy~ e^{-iyp} \tilde{f}(x,y)$~,  
\be 
\int\! dy ~e^{-iyp} \left(-{1\over 2m}
(\rp_y \pm{\hbar\over2} \rp_x)^2 + V(x\pm{\hbar\over2}y)-E\right)
\tilde{f} (x,y)=0.
\ee
Hence, Real solutions of (\ref{stargenvalue}) must 
be of the form 
\newline $f=\int\! dy~ e^{-iyp} 
\psi^*(x-{\hbar\over2}y) \psi(x+{\hbar\over2}y) /2\pi$,  
such that ${\mathfrak H}\psi =E\psi$. \qed 

The eqs (\ref{stargenvalue}) lead to 
 spectral properties  for WFs\cite{dbf,cfz}, as in the Hilbert space 
formulation. For instance, projective orthogonality of the $\*$-genfunctions
follows from associativity, which allows evaluation in two alternate groupings: 
\be
f\star H\star g= E_f ~f\star g =E_g ~f\star g.
\ee
Thus, for $E_g\neq E_f$,  it is necessary that 
\be f\star g= 0.
\ee 
Moreover, precluding degeneracy (which can be treated separately), choosing 
$f=g$ above yields, 
\be
f\star H\star f= E_f ~ f\star f= H\star f\star f ,
\ee
and hence $ f\star f$ must be the stargenfunction in question,
\be
f\star f \propto f. 
\ee
Pure state $f$s then $\star$-project onto their space. In general, 
it can be shown\cite{takabayasi,cfz} that 
\be  
{f_a\star f_b= \frac{1}{h}~ \delta_{a,b}~ f_a}.
\ee
The normalization matters\cite{takabayasi}: despite linearity of 
the equations, it prevents superposition of solutions. (Quantum mechanical 
interference works differently here, in comportance with density matrix 
formalism.)

By virtue of (\ref{Ndjambi}), for different $\star$-genfunctions, the 
above dictates that 
\be 
\int\! dp dx ~ fg  =0 ~.
\ee
Consequently, unless there is zero overlap for all such WFs,
at least one of the two must go negative someplace to offset the 
positive overlap\cite{reviews}---an illustration of the negative values' 
feature, quite far from being a liability.

Further note that integrating (\ref{stargenvalue}) yields the 
expectation of the energy, 
\be
\int\! H(x,p)f(x,p)~dxdp = E\int\! f~dxdp=E.
\ee

NB Likewise\footnote{
This discussion applies to proper WFs, corresponding to pure states' 
density matrices. E.g., a sum of two WFs is not a pure state in general, and 
does not satisfy the condition (\ref{test}). For such generalizations, the 
``impurity" is\cite{groen}~  $\int dx dp (f - h f^2) \geq 0$, where the 
inequality is only saturated into an equality for a pure state. 
For instance, for 
$w\equiv (f_a+f_b)/2$ with $f_a\* f_b=0$, the impurity is nonvanishing,
$\int dx dp ( w - h w^2)=1/2$.}, integrating the above projective 
condition yields
\be 
      \int\! f^2~dxdp = \frac{1}{h}~ ,
\ee 
that is the overlap increases to a divergent result in the classical limit,
as the WFs grow increasingly spiky.

In classical (non-negative) probability distribution theory, expectation 
values of non-negative functions are likewise non-negative, and thus result in 
standard constraint inequalities for the constituent pieces of such functions,
such as, e.g., moments of the variables. But it was just seen that for WFs
which go negative for an arbitrary function $g$, 
$\langle |g|^2 \rangle$ need not $\geq 0$: this can be easily seen by 
choosing the support of $g$ to lie
mostly in those regions of phase-space where the WF $f$ is negative. 

Still, such constraints are not lost for WFs. 
It turns out they are replaced by 
\be 
\langle g^* \star g \rangle \geq 0,
\ee
 and lead to the uncertainty principle\cite{uncertcz}.
In Hilbert space operator formalism, this relation
would correspond to the positivity of the norm.  
This expression is non-negative because it involves a real non-negative 
integrand for a pure state WF satisfying the above projective 
condition\footnote{ Similarly, if  $f_1$ and $f_2$ are pure state WFs, 
the transition probability between the respective states is also 
non-negative\cite{oconnwig}, manifestly  
by the same argument\cite{uncertcz}, namely, 
$\int\! dp dx f_1 f_2 = ({2\pi\hbar} )^2 \int   
dxdp\  |f_1 \star f_2 |^2 \geq 0.$},
\begin{equation}
\int\! dp dx (g^{*}\star g) f = 
h \int\!    dxdp (g^{*}\star g) (f\star f)  =
h\int\!    dxdp (f\star g^*)\star   (g\star f) = 
h\int\!    dxdp |g \star f |^2.
\end{equation}

To produce Heisenberg's 
uncertainty relation, one only need choose
\begin{equation}
g= a + bx+cp,
\end{equation}
for arbitrary complex coefficients $a,b,c$. The resulting positive 
semi-definite quadratic form is then
\begin{equation}
 a^*a  + b^* b \langle x \star x\rangle + c^* c \langle p\star p\rangle 
+(a^*b +b^* a) \langle x\rangle + (a^* c+c^* a) \langle p\rangle+ 
c^* b\langle p\star x \rangle+ b^* c \langle x\star p \rangle  \geq 0, 
\end{equation}
for any $a,b,c$.  The eigenvalues of the corresponding matrix are then 
non-negative, and thus so must be its determinant. Given 
\begin{equation}
x\star x=x^2, \qquad   p\star p=p^2, \qquad 
   p\star x= px  -i\hbar/2, \qquad  x\star p= px  +i\hbar/2, 
\end{equation}
and the usual 
\begin{equation}
(\Delta x )^2 \equiv  \langle (x-\langle x\rangle )^2 \rangle ,  
\qquad \qquad 
(\Delta p )^2 \equiv  \langle (p-\langle p\rangle )^2 \rangle ,  
\end{equation}
this condition on the $3\times 3$ matrix determinant amounts to
\begin{equation}
(\Delta x )^2 ~(\Delta p )^2 \geq \hbar^2/4 +\Bigl  (\langle 
(x-\langle x\rangle ) (p-\langle p\rangle) \rangle \Bigr  )^2, 
 \end{equation}
and hence 
\begin{equation}
\Delta x ~\Delta p \geq \hbar/2. 
\end{equation}
The $\hbar$ entered into the moments' constraint through the action of 
the $\*$-product.    \qed

\section{Illustration: the Harmonic Oscillator } 
\noindent To illustrate the formalism on a simple prototype 
problem, one may look 
at the harmonic oscillator. In the spirit of this picture, one can, in fact,
 eschew solving the Schr\"{o}dinger problem and plugging the wavefunctions into 
(\ref{wiggie}); instead, one may solve (\ref{stargenvalue}) directly for 
$H=(p^2 +x^2)/2$ (with $m=1$, $\omega=1$),
\be
\left( (x+{i\hbar \over 2} \partial_p)^2+(p-{i\hbar\over 2} \partial_x)^2 
-2E \right)  f(x,p)=0 .
\ee
For this hamiltonian, the equation has collapsed to two simple PDEs.
The first one, the Imaginary part,
\be (x\partial_p -p \partial_x)f =0,
\ee
restricts $f$ to depend on only one variable, the scalar in phase space,
$z=4H/\hbar=2(x^2+p^2)/\hbar$. Thus the second one, the Real part, is a 
simple ODE, 
\be
\left({z\over 4}-z\partial_z^2 -\partial_z -{E\over \hbar}\right) f(z) = 0.
\ee
Setting $f(z)=\exp(-z/2) L(z)$ yields Laguerre's equation, 
\be 
\left( z\partial_z^2 + (1-z)\partial_z +{E\over \hbar} -{1\over 2} 
\right) L(z) = 0.
\ee
It is solved by Laguerre polynomials,  
\be
L_n={e^z \partial^n (e^{-z} z^n)\over n! }  ~,
\ee
for $n=E/\hbar -1/2=0,1,2,..$, so the $\*$-gen-Wigner-functions are 
\cite{groen}
\be
&&f_n= \frac{(-1)^n}{\pi}~ e^{-2H/\hbar} ~L_n(4H/\hbar); \nonumber \\
&&L_0= 1,
\quad  L_1=1-4H/\hbar, \quad  
L_2=8H^2/\hbar^2-8 H/\hbar +1, ...
\ee
But for the Gaussian ground state, they all have zeros and 
go negative. These functions become spiky in the classical limit 
$\hbar\rightarrow 0$; e.g., the ground state Gaussian $f_0$ goes to a 
$\delta$-function. 

Their sum provides a resolution of the identity\cite{moyal}, 
\be
\sum_n f_n = 1/h .
\ee

(For the rest of this section, for algebraic simplicity, set $\hbar=1$.)

\begin{figure}[htbp] %ORIGINAL SIZE: width=4TRUEIN; height=3TRUEIN
\vspace*{13pt}
\centerline{\psfig{file=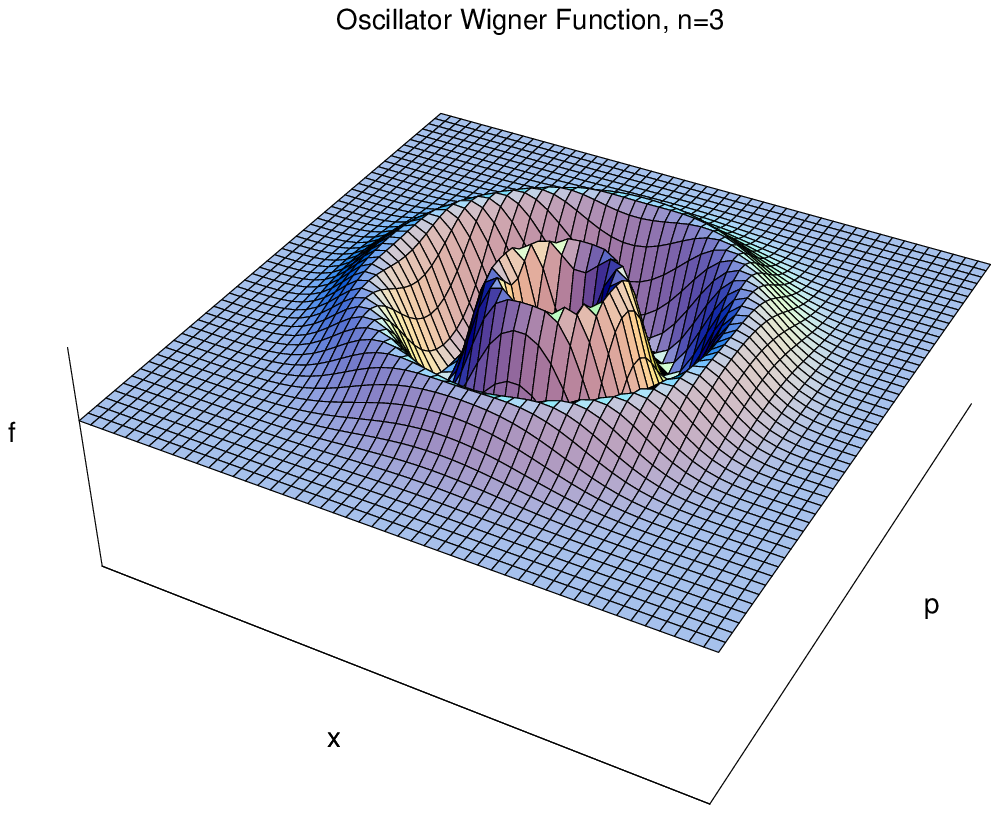}} %90 percent
\vspace*{13pt}
\fcaption{WF for the 3rd excited state. Note the negative values.}
\end{figure}

Dirac's hamiltonian factorization method for the alternate algebraic solution 
of the same problem carries through intact, with $\star$-multiplication 
supplanting operator multiplication. That is to say,
\be
H= \frac{1}{2} (x-ip) \star (x+ip) + \frac{1}{2} ~.
\ee
This motivates definition of raising and lowering functions (not operators) 
\be 
a\equiv \frac{1}{\sqrt{2}} (x+ip),   \qquad \qquad 
a^{\dagger} \equiv \frac{1}{\sqrt{2}} (x-ip),
\ee
where 
\be
a \star     a^{\dagger} -a^{\dagger} \star a=1 ~.  
\ee
The annihilation ones $\*$-annihilate the $\star$-Fock vacuum,   
\be
a \star f_0= \frac{1}{\sqrt{2}} (x+ip) 
\star e^{-(x^2+p^2)}=0~.
\ee

Thus, the associativity 
of the $\star$-product permits the customary ladder spectrum 
generation\cite{cfz}. The $\star$-genstates for $H\star f= f\star H$ are 
\be 
{f_n\propto (a^{\dagger}\star)^n   ~ f_0 ~(\star a)^n ~.  }
\ee
They are manifestly real, like the Gaussian ground state,
and left-right symmetric; it is easy to see they are $\star$-orthogonal 
for different eigenvalues. Likewise, they can be seen by the evident algebraic 
normal ordering to project to themselves, since the Gaussian ground 
state does, $f_0 \star f_0 = f_0/h$.   The corresponding coherent state WFs 
\cite{kim,cuz} are likewise very analogous to the conventional formulation.

This type of analysis carries over well to a broader class of 
problems\cite{cfz} with ``essentially isospectral" pairs of partner potentials,
connected with each other through Darboux transformations relying on Witten 
superpotentials $W$ (cf the P\"{o}schl-Teller potential). It closely 
parallels the standard differential equation structure of the recursive 
technique. That is, the pairs of related potentials and corresponding 
$\*$-genstate Wigner functions are constructed recursively\cite{cfz} 
through ladder operations analogous to the algebraic method 
outlined for the oscillator. 

Beyond such recursive potentials, examples of further simple systems where 
the $\*$-genvalue equations can be 
solved on first principles include the linear potential 
\cite{garcia,cfz,torres}, 
the exponential 
interaction Liouville potentials, and their supersymmetric Morse 
generalizations\cite {cfz} (also see \cite{frank}). 
Further systems may be handled through the Chebyshev-polynomial 
numerical techniques of ref \cite{hug}.
 
First principles phase-space solution of the Hydrogen atom is less than 
straightforward and complete. The reader is referred to 
\cite{bayen,bondia,dspringborg} for significant partial results.

\section {Time Evolution }
\noindent Moyal's equation (\ref{evolution}) is formally solved by virtue of 
associative combinatoric 
operations completely analogous to Hilbert space Quantum Mechanics, through 
definition of a $\star$-unitary evolution operator, 
a ``$\star$-exponential"\cite{bayen} 
\be 
U_{\star} (x,p;t)=e_\star ^{itH/\hbar} \equiv 
1+(it/\hbar)H(x,p) + {(it/\hbar)^2\over 2!}  H\star H +{(it/\hbar )^3\over 3!} 
 H\star H\star H +...,
\ee 
for arbitrary hamiltonians. The solution to Moyal's equation,
given the WF at $t=0$, then, is 
\be
f(x,p;t)=U_{\star}^{-1} (x,p;t) \star f(x,p;0)\star U_{\star}(x,p;t) .    
\ee 

For the variables $x$ and $p$, the evolution equations 
collapse to {\em classical} trajectories,
\be
{d x \over dt}={ x \star H-H\star x \over  i\hbar}= \partial_p H = p ~ ,
\ee
\be 
{d p \over dt}={ p \star H-H\star p \over i\hbar}=- \partial_x H =-x~,
\ee
where the concluding member of these two equations hold 
for the oscillator only. Thus, for 
the oscillator, 
\be 
x(t)= x \cos t + p \sin t,  
\ee
\be
p(t)=p \cos t - x \sin t.     
\ee

As a consequence, for the oscillator, the functional form of the Wigner 
function is 
preserved along classical phase-space trajectories\cite{groen}, 
\be 
f(x,p;t)=f(x \cos t - p \sin t,     p \cos t + x \sin t  ;0). 
\ee

\begin{figure}[htbp] 
\vspace*{6pt}
\centerline{\psfig{file=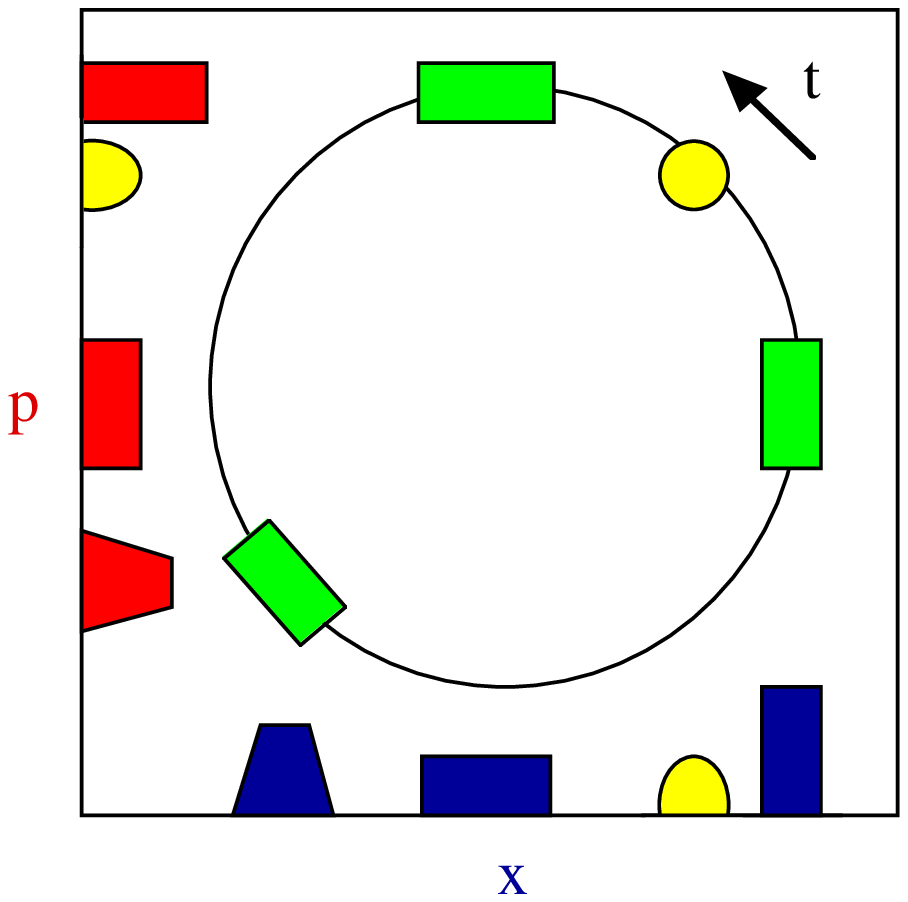}} 
\vspace*{6pt}
\fcaption{
 Time evolution of generic WF configurations driven by an oscillator 
hamiltonian. The $ t$-arrow indicates the rotation sense of $x$ and $p$, and
so, for fixed $x$ and $p$ axes, the WF shoebox configurations rotate rigidly 
in the opposite direction, clockwise. (The sharp angles of the WFs 
in the illustration are actually unphysical, and were only chosen to 
monitor their 
``spreading wavepacket" projections more conspicuously.) These $x$ and 
$p$-projections (shadows) are meant to be
intensity profiles on those axes, but are expanded on the plane to aid
visualization. The circular figure represents a coherent state, which 
projects on either axis identically at all times, thus without 
shape alteration of its wavepacket through time evolution.}
\end{figure}

{\em Any} oscillator WF configuration rotates 
uniformly on the phase plane around the origin, essentially classically,
even though it provides a complete quantum mechanical 
description\cite{groen,bartlettMoyal,wigner,lesche,base,kyotozc}.

Naturally, this rigid rotation in phase-space preserves areas, and
thus automatically illustrates the uncertainty principle. 
By contrast, in general, in the 
conventional formulation of quantum mechanics, this result is deprived of 
intuitive import, or, at the very least, simplicity: upon integration in $x$ 
(or $p$) to yield usual probability densities, the rotation induces 
apparent complicated shape variations of the oscillating probability density 
profile, such as wavepacket spreading (as evident in the shadow 
projections on the $x$ and $p$ axes of Fig 2). 

Only when 
(as is the case for coherent states\cite{hennings}) a Wigner function 
configuration has an 
{\em additional} axial $x-p$ symmetry around its {\em own} center, 
will it possess an invariant profile upon this rotation, and hence a 
shape-invariant oscillating probability density\cite{kyotozc}. 

In Dirac's interaction representation, a more
complicated interaction hamiltonian superposed on the oscillator one, 
leads to shape changes 
of the WF configurations placed on the above ``turntable", and serves to 
generalize to scalar field theory\cite{base}.
A more elaborate discussion of propagators in found in \cite{cuz,garcia}.
\pagebreak

\section{Canonical Transformations }
\noindent
Canonical transformations $(x,p)\!\mapsto\! (X(x,p), P(x,p))$ 
preserve the phase-space volume (area) element (again, take $\hbar=1$) through a trivial 
Jacobian, 
\be 
dXdP= dx dp ~\{  X, P  \}~,
\ee
ie, they preserve Poisson Brackets 
\be
\{u,v \}_{xp}\equiv {\partial   u \over \partial x  }
{\partial v    \over \partial p }-
{\partial u    \over \partial p  }
{\partial v   \over \partial x }~,
\ee
\be
\{  X, P  \}_{xp} = 1  ,\qquad \{  x, p  \}_{XP} = 1 ,\qquad  
\{  x, p  \}=\{  X, P  \}.
\ee

Upon quantization, the c-number function hamiltonian transforms 
``classically", ${\cal H}(X,P)\equiv H(x,p)$, like a scalar.
Does the $\star$-product remain invariant under this transformation?

Yes, for {\em linear} canonical transformations\cite{kim}, but clearly 
{\em not for general canonical transformations}. 
Still, things can be put right, by devising
general {\em covariant} transformation rules for the 
$\star$-product\cite{cfz}: The WF transforms in comportance with 
Dirac's quantum canonical transformation theory\cite{pamd}.
  
In conventional quantum mechanics, for canonical transformations generated 
by $F(x,X)$,
\be
p={\partial F(x,X) \over \partial x } ~, \qquad 
P=-{\partial F(x,X) \over \partial X }~, 
\ee
the energy eigenfunctions transform in a generalization of the 
``representation-changing" Fourier transform\cite{pamd},  
\be
\psi _{E}(x)=N_{E}\int dX\,e^{iF(x,X)}\,\Psi _{E}(X) ~.
\ee
(In this expression, the generating function $F$ may contain $\hbar$ 
corrections to the classical one, in general---but for several simple quantum 
mechanical systems it manages not to\cite{ghandour}.)
Hence\cite{cfz}, there is a transformation functional for WFs,
${\cal T}(x,p;X,P)$, such that  
\be
f(x,p)=\int \!dX dP\;{\cal T} (x,p;X,P)
\;\bigcirc\kern-0.9em\star\; {\cal F}(X,P)                  
=\int \!dX dP\;{\cal T} (x,p;X,P)\; {\cal F}(X,P)\;, 
\ee 
where 
\be 
&& {\cal T}(x,p;X,P)    \\
&& = \frac{|N|^2}{2\pi} \!\int\!dY dy~ \exp \left ( -iyp+iPY 
-iF^{*}(x-{\frac{y}{2}},X-{\frac{Y}{2}}) 
+iF(x+{\frac{y}{2}} ,X+{\frac{Y}{2}})\right ). \nonumber  
\ee
Moreover, it can be shown that\cite{cfz},
\be 
H(x,p)\star {\cal T}(x,p;X,P) =
{\cal T}(x,p;X,P) \bigcirc \kern-0.9em\star ~ {\cal H}(X,P).   
\ee
That is, if ${\cal F}$ satisfies a $\;\bigcirc \kern-0.9em\star-$genvalue
equation, then $f$ satisfies a $\star $-genvalue equation with 
the same eigenvalue, and vice versa. This proves useful in 
constructing WFs for simple systems which can be trivialized classically 
through canonical transformations. 

\section{Omitted Miscellany}
\noindent Phase-space quantization extends in several interesting 
directions which are not covered in such a short introduction. 

Complete sets of WFs which correspond to {\em off-diagonal} elements of the 
density matrix (see next section), and which thus enable investigation of 
interference phenomena and the  formulation of quantum mechanical 
perturbation theory are covered in \cite{bartlettMoyal,wangoconnell,cuz}.
The spectral theory of WFs is discussed in \cite{bayen,manogue,cuz}.
Spin is treated in ref \cite{varilly}.

Mathematical, uniqueness,  and computational
features of $\*$-products in various geometries are discussed in 
\cite{bayen,hansen,triangle,kontsevich,wakatsuki,vvoros,estrada,antonsen}.

Alternate phase-space distributions, but always equivalent to the WF, are
reviewed in \cite{reviews,voros,triangle,takahashi}.  

\section{The Weyl Correspondence}
\noindent 
This section summarizes the bridge and equivalence of phase-space quantization 
to the conventional formulation of Quantum Mechanics in Hilbert space. 

Weyl\cite{weyl} introduced an association rule 
mapping invertibly  c-number phase-space 
functions $g(x,p)$ (called classical kernels) to operators ${\mathfrak G}$ in 
a given ordering prescription. Specifically, 
$p\mapsto{\mathfrak p}$, $x\mapsto{\mathfrak x}$, and, in general,
\begin{equation}      
{\mathfrak G}({\mathfrak x},{\mathfrak p}) 
=\frac{1}{(2\pi)^2}\int d\tau d\sigma 
dx dp ~g(x,p) \exp (i\tau ({\mathfrak p}-p)+i\sigma ({\mathfrak x}-x)).
\end{equation}
The eponymous ordering prescription requires that an arbitrary operator, 
regarded as a power series in ${\mathfrak x}$ and ${\mathfrak p}$, be first 
ordered in a completely symmetrized expression in ${\mathfrak x}$ and 
${\mathfrak p}$, 
by use of Heisenberg's commutation relations, $ [{\mathfrak x},{\mathfrak p} ]
 =i\hbar $. A term with 
$m$ powers of ${\mathfrak p}$ and $n$ powers of ${\mathfrak x}$ will be 
obtained from the coefficient of $\tau^m \sigma^n$ in the expansion of 
$(\tau {\mathfrak p} + \sigma {\mathfrak x} )^{m+n}$. 
It is evident how the map yields a Weyl-ordered operator from a polynomial 
classical kernel. It includes every possible ordering with multiplicity one, 
eg, 
\begin{equation}   
6 p^2 x^2 \mapsto {\mathfrak p}^2 {\mathfrak x}^2 + 
 {\mathfrak x}^2  {\mathfrak p}^2 + 
 {\mathfrak p}  {\mathfrak x}  {\mathfrak p}  {\mathfrak x} + 
 {\mathfrak p}  {\mathfrak x}^2  {\mathfrak p} + 
 {\mathfrak x}  {\mathfrak p}  {\mathfrak x}  {\mathfrak p} + 
 {\mathfrak x}  {\mathfrak p}^2  {\mathfrak x}~.
\end{equation}

In this correspondence scheme, then, 
\be
\hbox{Tr}{\mathfrak G} = \int dx dp~ g.
\ee

Conversely\cite{groen}, the c-number phase-space kernels $g(x,p)$
of Weyl-ordered operators ${\mathfrak G}({\mathfrak x},{\mathfrak p})$ 
are specified by  
${\mathfrak p} \mapsto p$, ${\mathfrak x}\mapsto x $, or, 
more precisely,   
\be
g(x,p) &=&\frac{1}{(2\pi)^2} \int d\tau  d\sigma ~ e^{i(\tau p + \sigma x)} 
\hbox{Tr~}\left ( 
e^{-i(\tau {\mathfrak p} + \sigma {\mathfrak x})} {\mathfrak G} \right )   
\nonumber \\ 
&=&\frac{1}{2\pi} \int dy~ e^{-iyp} \langle x +\frac{\hbar}{2}y 
| {\mathfrak G}({\mathfrak x},{\mathfrak p}) | x-\frac{\hbar}{2}y \rangle ,
\ee 
since the above trace reduces to 
\be
\int dz ~ e^{i\tau  \sigma \hbar /2 }  \langle z | 
e^{-i\sigma {\mathfrak x}}    e^{-i \tau {\mathfrak p}} {\mathfrak G} 
| z \rangle = {2\pi} 
\int dz \langle z -\hbar \tau | {\mathfrak G} | z \rangle 
 e^{i\sigma (\tau  \hbar /2- z) }  . 
\ee

Thus, the density matrix inserted in this expression\cite{moyal} yields 
a hermitean generalization of the Wigner function:
\be
f_{mn}(x,p)&\equiv &\frac{1}{2\pi}\int dy~ e^{-iyp} \langle x+\frac{\hbar}{2}y 
| \psi_n \rangle \langle \psi_m | x-\frac{\hbar}{2}y \rangle  \nonumber\\
&=& \frac{1}{2\pi}\int dy e^{-iyp} \psi^*_m(x-\frac{\hbar}{2}y)
\psi_n(x+\frac{\hbar}{2}y)=f_{nm}^* (x,p)~,   
\ee
where the $\psi_m (x)$s are (ortho-)normalized solutions of a 
Schr\"{o}dinger problem. (Wigner\cite{wigner} mainly considered the diagonal 
elements of the density matrix 
(pure states), denoted above as $f_m\equiv f_{mm}$.)  
As a consequence, matrix elements of operators, ie, traces of them with 
the density matrix, are produced through mere phase-space 
integrals\cite{moyal},
\be
\langle \psi_m | {\mathfrak G} |\psi_n \rangle =
\int dx dp ~g(x,p) f_{mn}(x,p),   \label{mamiwata}
\ee
and thus expectation values follow for $m=n$, as utilized 
throughout in this review. Hence, 
\be
\langle \psi_m |\exp{i(\sigma {\mathfrak x}+\tau {\mathfrak p}  }  )
 |\psi_m \rangle =
\int dx dp ~ f_{m}(x,p) \exp{i(\sigma x + \tau p)} ,
\ee
the moment-generating functional\cite{moyal} of the Wigner distribution. 

Products of Weyl-ordered operators are easily reordered into 
Weyl-ordered operators through the degenerate Campbell-Baker-Hausdorff identity.
In a study of the uniqueness of the Schr\"{o}dinger representation,
von Neumann\cite{neumann} adumbrated the composition rule of classical 
kernels in such operator products, appreciating that Weyl's correspondence 
was in fact a homomorphism. (Effectively, he arrived at a convolution 
representation of the star product.) Finally, Groenewold\cite{groen} 
neatly worked out in detail how the classical 
kernels $f$ and $g$ of two operators ${\mathfrak F}$ and ${\mathfrak G}$ 
must compose to yield the classical kernel of ${\mathfrak F} {\mathfrak G}$,
\begin{eqnarray}  
{\mathfrak F}{\mathfrak G}&=& 
 \frac{1}{(2\pi)^4} \int {d\xi d\eta d\xi' d\eta'
dx' dx'' dp' dp''}   f(x',p') g(x'',p'')\nonumber \\
&\times  &\exp i(\xi ( {\mathfrak p}-p')+\eta ( {\mathfrak x}-x'))~
\exp i(\xi' ( {\mathfrak p}-p'')+\eta' ( {\mathfrak x}-x''))=  \nonumber  
\end{eqnarray}  
$$
 =\frac{1}{(2\pi)^4}\!\!\int\! d\xi d\eta d\xi' d\eta' 
dx' dx'' dp' dp'' f(x',p') g(x'',p'') 
\exp i\left( (\xi +\xi')  {\mathfrak p}+(\eta+\eta')  {\mathfrak x}\right) 
$$
\be
\times ~\exp i\!\left(-\xi p'-\eta x'-\xi' 
p''-\eta'x'' +{\hbar\over 2} (\xi\eta'-\eta\xi') \right).
\ee
Changing integration variables to
\begin{equation}  
\xi'\equiv {2\over \hbar} (x-x'), \quad
\xi\equiv \tau-  {2\over \hbar} (x-x'), \quad
\eta'\equiv {2\over \hbar} (p'-p), \quad
\eta\equiv \sigma- {2\over \hbar} (p'-p), 
\end{equation}
reduces the above integral to
\begin{equation}
 {\mathfrak F}{\mathfrak G}= \frac{1}{(2\pi)^2}\int d\tau d\sigma 
dx dp ~\exp i \left(\tau ( {\mathfrak p}-p)+\sigma ( {\mathfrak x}-x)\right)
~(f\star g)(x,p), 
\end{equation}
where $f\star g$ is the expression (\ref{fourier}). \qed

\nonumsection{Acknowledgements}
\noindent
I thank D Fairlie and T Curtright for helpful comments.
This work was supported in part by the US Department of Energy, 
Division of High Energy Physics, Contract W-31-109-ENG-38.

\nonumsection{References} 

\phantom{aa}

\phantom{aa}

\appendix{\qquad Selected Publications on Phase-space Quantization}
\noindent
Implicitly, the bulk of the formulation is contained in Groenewold's and 
Moyal's seminal papers. But this has been a slow story of emerging connections 
and chains of ever sharper reformulations, and confidence in the autonomy of
the formulation arose very slowly. As a result, attribution of critical 
milestones cannot avoid subjectivity: it cannot automatically highlight 
merely the {\em earliest}
occurrence of a construct, unless that has also been compelling enough to yield
an ``indefinite stay against confusion"  about the logical structure of the 
formulation.

H Weyl (1927)\cite{weyl} 
introduces the correspondence of ``Weyl-ordered" operators to phase-space 
(c-number) kernel functions (as well as discrete QM application of 
Sylvester's (1883)\cite{sylvester} clock and shift matrices).
  
J von Neumann (1931)\cite{neumann},
 in a technical aside off a study of the uniqueness of Schr\"{o}dinger's 
representation, includes an implicit version of the $\star$-product 
which promotes Weyl's correspondence rule to full isomorphism between 
Weyl-ordered operator multiplication and $\star$-convolution of kernel 
functions.

E Wigner (1932)\cite{wigner} (and L Szilard, unpublished) 
introduces the eponymous phase-space distribution function controlling quantum
mechanical diffusive flow in phase space. It specifies the time evolution of
this function and applies it to quantum statistical mechanics.

H Groenewold (1946)\cite{groen}.  
A seminal but somewhat unappreciated paper which 
fully understands the Weyl correspondence and produces the WF as the 
classical kernel of the density matrix. It reinvents and streamlines 
von Neumann's construct into the standard 
$\star$-product, in a systematic exploration of the isomorphism 
between Weyl-ordered operator products and their kernel function compositions. 
It further works out the harmonic oscillator WF.

J Moyal (1949)\cite{moyal} amounts to a grand  synthesis: 
It establishes an independent formulation 
of quantum mechanics in phase space. It systematically studies 
all expectation values of Weyl-ordered operators, and identifies 
the Fourier transform of their moment-generating function (their 
characteristic function) to the Wigner Function.  
It further interprets the subtlety of the ``negative probability" formalism
and reconciles it with the uncertainty principle. Not least, it recasts the 
time evolution of the Wigner Function through a deformation of the Poisson 
Bracket into the Moyal Bracket (the commutator of $\star$-products, i.e.,
the Weyl correspondent of the Heisenberg commutator),
and thus opens up the way for a systematic study of the semiclassical limit. 
Before publication, Dirac has already been impressed by this 
work, contrasting it to his own ideas on functional integration, in Bohr's
Festschrift\cite{pamdbohr}.

T Takabayasi (1954)\cite{takabayasi} 
investigates the fundamental projective normalization condition for pure state 
Wigner functions, and exploits Groenewold's link to the conventional density 
matrix formulation. It further illuminates 
the diffusion of wavepackets. 

G Baker (1958)\cite{baker} 
envisions the logical autonomy of the formulation, based on postulating 
the projective normalization condition. It resolves measurement subtleties in 
the correspondence principle and appreciates the significance of the 
anticommutator of the $\star$-product as well, thus shifting emphasis to 
the $\star$-product itself, over and above its commutator.

D Fairlie (1964)\cite{dbf} (also see 
\cite{starothers}) 
explores the time-independent counterpart to Moyal's evolution equation, 
which involves the $\star$-product, beyond mere Moyal Bracket equations, and 
derives (instead of postulating) the projective orthonormality 
conditions for the resulting Wigner functions. These now 
allow for a unique and full solution  of the quantum system, in principle
(without any reference to the conventional Hilbert-space formulation).
Thus, autonomy of the formulation is fully recognized.

M Berry (1977)\cite{berry}
elucidates the subtleties of the semiclassical limit, ergodicity,
integrability, and the singularity structure of Wigner function evolution.

F Bayen, M Flato, C Fronsdal, A Lichnerowicz, and D Sternheimer (1978) 
\cite{bayen} 
analyzes systematically the deformation structure and the uniqueness
of the formulation, with special emphasis on spectral theory, 
and consolidates it mathematically. It provides explicit
solutions to standard problems and introduces influential technical 
innovations, such as the $\star$-exponential. 

T Curtright, D Fairlie, and C Zachos (1998)\cite{cfz} 
demonstrates more directly the equivalence of the time-independent 
$\star$-genvalue problem to the Hilbert space formulation, and hence its 
logical autonomy; formulates Darboux isospectral systems in phase space; 
establishes the covariant  transformation 
rule for general {\em nonlinear} canonical transformations (with reliance on 
the classic work of P Dirac (1933)\cite{pamd}; 
and thus furnishes explicit solutions of nontrivial practical problems on 
first principles, without recourse to the Hilbert space formulation. Efficient 
techniques, e.g. for perturbation theory, are based on generating functions
for complete sets of Wigner functions in T Curtright, T Uematsu, 
and C Zachos (2001)\cite{cuz}. A self-contained derivation of the uncertainty 
principle is provided in T Curtright and C Zachos (2001)\cite{uncertcz}. 

M Hug, C Menke, and W Schleich (1998)\cite {hug} 
introduce techniques for numerical solution of $\star$-equations on a basis of 
Chebyshev polynomials.
\end{document}